\documentclass[%
reprint,
amsmath,amssymb,
aps,
]{revtex4-2}

\usepackage{amsmath}
\usepackage{graphicx}% Include figure files
\usepackage{dcolumn}% Align table columns on decimal point
\usepackage{bm}% bold math
\usepackage[colorlinks=true, citecolor=blue, linkcolor=blue]{hyperref}

\usepackage{xcolor}
\usepackage{subfigure}

\renewcommand{\textcolor}[2]{#2}

\begin{document}
	
	\preprint{APS/123-QED}
	
	\title{{Nonlinear Dynamics of Hopfion for Frequency Multiplication}}% Force line breaks with \\
	
	\author{Waleed I. Waseer}
	\affiliation{School of Physics and State Key Laboratory of Electronics Thin Films and Integrated Devices, University of Electronics and Science and Technology of China, Chengdu 610054, China}

%	\author{Tijjani Abdulrazak}
%\affiliation{School of Physics and State Key Laboratory of Electronics Thin Films and Integrated Devices, University of Electronics and Science and Technology of China, Chengdu 610054, China}

\author{Yunshan Cao}
\affiliation{School of Physics and State Key Laboratory of Electronics Thin Films and Integrated Devices, University of Electronics and Science and Technology of China, Chengdu 610054, China}
	\author{Peng Yan}%
	\email{yan@uestc.edu.cn}
\affiliation{School of Physics and State Key Laboratory of Electronics Thin Films and Integrated Devices, University of Electronics and Science and Technology of China, Chengdu 610054, China}
	
%	\altaffiliation{School of Physics and State Key Laboratory of Electronics Thin Films and integrated devices, University of Electronics and Science and Technology of China, Chengdu 610054}% 

	\begin{abstract}
		Hopfions, associated with higher-dimensional topology through the Hopf fibration, exhibit {exotic} features like {complex knot} and improved stability compared to skyrmions, enhancing their appeal for innovative applications. In this paper, we study the nonlinear response of magnetic Hopfion to microwave fields. {We observe the emergence of higher-order harmonics of the driving microwave field as it interacts with the Hopfion}. By carefully selecting the {driving frequency, the corresponding harmonic can efficiently excite localized magnon state of a Hopfion. Our results demonstrate the promising potential of hopfions in nonlinear magnonics. }

%		applied frequency as a fraction of the resonance frequency of the localized magnon state, the corresponding harmonics can be utilized to excite localized magnon states of a Hopfion. In the case of skyrmions, these localized magnon states have demonstrated potential for various applications, including racetrack memories, microwave generation, and unconventional computing. Exciting a localized state through its corresponding harmonics—which, unlike eigenmodes, are not intrinsic solutions of the system and may decay more rapidly—can be more efficient at specific field amplitudes, making the experimental excitation and detection of these localized modes more feasible. Furthermore, we demonstrate that both in-plane and out-of-plane  microwave filed can excite bounded magnon modes. Our results highlights that selecting an appropriate applied field strength and damping coefficient can induce the first, second, third, or fourth harmonics, leading to the excitation of magon states bounded to the hopfion.		  
	\end{abstract}
	
	
	\maketitle

\section{Introduction}
Topological solitons are stable, particle-like field configurations that cannot be continuously deformed to a uniform state due to their non-trivial integer topological index \cite{manton2004topological,nitta2022relations,sutcliffe2018hopfions}. This inherent topological protection makes them exceptionally robust against external perturbations. A celebrated example is the magnetic skyrmion, a two-dimensional soliton in nanoscale chiral magnets \cite{skyrme1962unified}. Skyrmions exhibit unique stability and dynamics, and the remarkably low currents required to manipulate them make them prime candidates for energy-efficient data storage and logic technologies \cite{yu2010real,melcher2014chiral,bogdanov1989thermodynamically,fert2013skyrmions,kiselev2011chiral}.

Despite the success of low-dimensional topological solitons, their high-dimensional counterparts remain less explored. A natural three-dimensional (3D) generalization of a skyrmion is the Hopfion, a soliton characterized by a closed-loop, twisted skyrmion string structure \cite{sutcliffe2018hopfions}. Its topological charge is defined by the Hopf index. A deep understanding of Hopfion statics and dynamics is of significant fundamental interest and may unlock novel applications.
\textcolor{red}{First theorized decades ago \cite{faddeev1997stable,kosevich1990magnetic}, magnetic Hopfions have only recently been observed by experiments \cite{ackerman2017static,kent2021creation}. Subsequent work demonstrated their stabilization in confined nanostructures and the realization of three-dimensional bulk Hopfion states in chiral magnetic systems \cite{zheng2023hopfion}.} They have also been studied across various fields, including liquid crystals, cosmology, and quantum systems \cite{faddeev1999partially,cooper1999propagating,babaev2002dual,babaev2002hidden,kleckner2013creation,ackerman2017static}. Dynamical studies have revealed intriguing phenomena like magnon focusing \cite{saji2023hopfion} and novel dynamics, including three-dimensional gyration, complex transformations, and knotting/un-knotting motions \cite{liu2018binding,raftrey2021field,wang2019current,khodzhaev2022hopfion,zheng2023hopfion}. This richer dynamics stems from their higher dimensionality and complex topology, which is linked to the Hopf fibration, allowing for features like knotting in four-dimensional space and contributing to their enhanced stability compared to skyrmions \cite{kent2021creation,guslienko2024magnetic}.

The excitation and control of magnetic textures often rely on spin waves, or magnons—the bosonic quasiparticles representing the precession of magnetic moments. Their classical dynamics are governed by the inherently nonlinear Landau-Lifshitz-Gilbert (LLG) equation \cite{kruglyak2010magnonics,gilbert2004phenomenological,lenk2011building}. This nonlinearity enables crucial effects such as frequency multiplication (e.g., doubling or tripling) \cite{demidov2011generation,sebastian2013nonlinear,rousseau2014propagation} and complex magnon scattering \cite{demidov2011generation,heinrich1985fmr,lenz2006two,fleury1968scattering,hurben1998theory,zakeri2007spin}. For applications, nonlinear frequency multiplication is particularly attractive as a means to overcome the limitations of conventional microwave antennas, where the excitation efficiency is restricted by the wavelength and sample size \cite{gross2022imaging}.

A key aspect of topological textures is their rich set of localized magnon modes including breathing, gyration, and higher-order excitations which are confined to the soliton core and play a pivotal role in applications such as microwave generation and unconventional computing~\cite{rodrigues2021nonlinear,dussaux2010large,nishimura2002magnetic,tomasello2014strategy,carpentieri2015topological,ruotolo2009phase}. These modes possess discrete eigenfrequencies. An efficient strategy to excite them is through nonlinear harmonic generation, whereby the system is driven at a subharmonic of the mode frequency (e.g., $f_{\text{drive}} = f_{\text{mode}}/2$), eliciting a strong response at the fundamental ($f_{\text{mode}}$). In contrast to parametric excitation \cite{zakharov1975spin,bryant1988spin,xiao2017parametric}, this approach exploits the soliton’s topology in a complementary fashion. It offers several advantages: the low-frequency driving field decays rapidly outside the localized texture, suppressing radiative losses and mitigating instabilities; it enables the use of lower-frequency, more practical microwave sources to access high-frequency responses; and it leverages intrinsic nonlinearities to produce a pronounced harmonic output at relatively low driving amplitudes \cite{rodrigues2021nonlinear}.

In this context, Hopfions offer additional opportunities for nonlinear dynamics. Their higher-dimensional topology, described by the Hopf fibration, allows for knotted and linked spin configurations that stabilize complex localized states beyond those in skyrmions \cite{kent2021creation,zheng2023hopfion}. These topological features enable enhanced mode coupling, so nonlinear responses such as harmonic generation can be achieved at relatively low driving amplitudes compared to uniform ferromagnets or systems without Hopfions. This robustness and efficiency position Hopfions as promising platforms for nonlinear magnonics and frequency-multiplication technologies \cite{guslienko2024magnetic}.

In this work, we show that the topologically localized potential of a Hopfion provides a natural and highly efficient platform for nonlinear frequency multiplication. Using systematic micromagnetic simulations in MuMax3, we excite a Hopfion in a chiral ferromagnetic nanodisk via microwave fields. The resulting power spectra exhibit pronounced second- and third-harmonic generation at relatively low driving amplitudes, indicating that the required input power is substantially lower than that needed to achieve comparable nonlinear effects in unstructured magnetic films or via direct high-frequency excitation. Importantly, we demonstrate that tuning the drive frequency to a fractional resonance of a Hopfion-localized mode (e.g., $f_{\text{drive}} = f_{\text{mode}}/2$) significantly enhances the harmonic response. These results uncover a topology-driven mechanism for frequency multiplication that exploits the three-dimensional confinement of Hopfions, highlighting their promise as compact, low-power magnonic frequency multipliers and signal processors for future spintronic applications.

The article is organized as follows: In Sec. II, we detail the simulation methods for stabilizing a Hopfion and characterizing its localized magnon modes. In Sec. III, we explore the field-driven nonlinear dynamics, providing multiple examples where localized states are efficiently excited at a fraction of their eigenfrequency. Conclusions are presented in Sec. IV.

\begin{figure}
		%\centering
		%\begin{minipage}{\linewidth}  % Adjust width as needed (0.7 is 70% of page width)
		%  \centering
		\includegraphics[width=\linewidth]{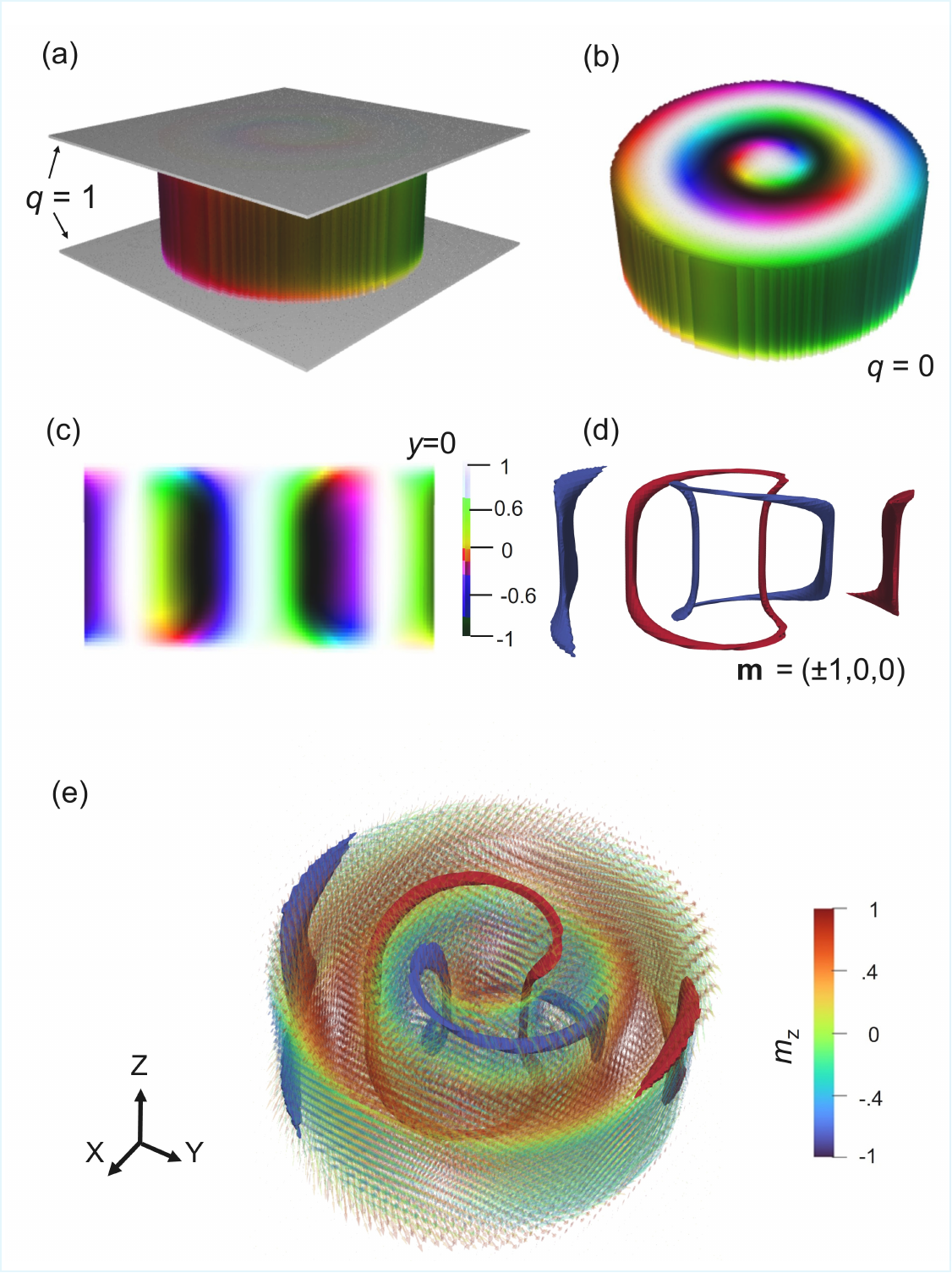}
		\caption{ {(a)} Schematic of a Hopfion stabilized in a chiral ferromagnetic disk sandwiched between two PMA layers. {(b)} $m_z$ component of a numerically computed Hopfion for $D=395$ $\mu$J/m$^2$ in a cylinder of height $h$ and diameter of $200.2$ nm. {(c)} Cross section of Hopfion at $y=0$. {(d)} The two isosurface at $\bold{m}=\left(\pm1,0,0\right)$  are linked once, as required by the unit Hopf charge. {(e)} Plot of  $m_\text{z}$ with two isosurface drawn $\bold{m}=\left(\pm1,0,0\right)$ representing the Hopf charge.  }
		\label{fig1}
		% \end{minipage}
\end{figure}

\begin{figure*}[t!]
	\centering
	\includegraphics[width=.95\linewidth]{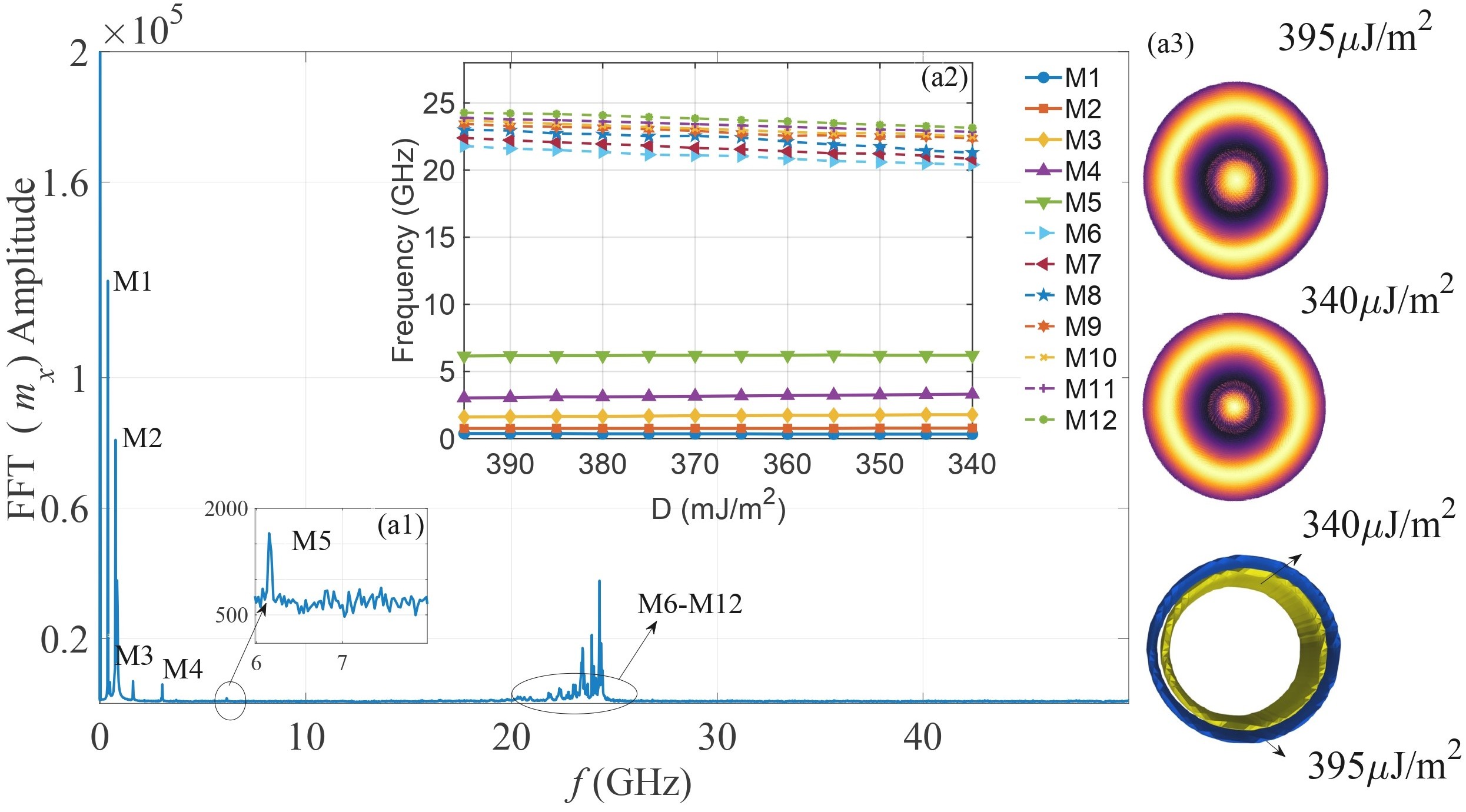 }
	\caption{\textcolor{red}{Spin-wave resonance spectrum {in response to an in-plane excitation}. Inset (a1) shows a magnified view of the magnon spectrum, highlighting the labeled localized state M5. The inset (a2) illustrates the behavior of localized states in response to DMI. Modes M6-M12 corresponds to 21.8,22.4 ,23,23.45,23.7,23.9 and 24.275 GHz respectively.  The inset (a3) illustrates the effect of varying DMI on the Hopfion profile along with the contour drawn at $ \mathbf{m}=(0,0,-1)$.}}
	\label{fig2}
\end{figure*} 
\begin{figure}[hbt!]
	\centering
	\includegraphics[width=\linewidth]{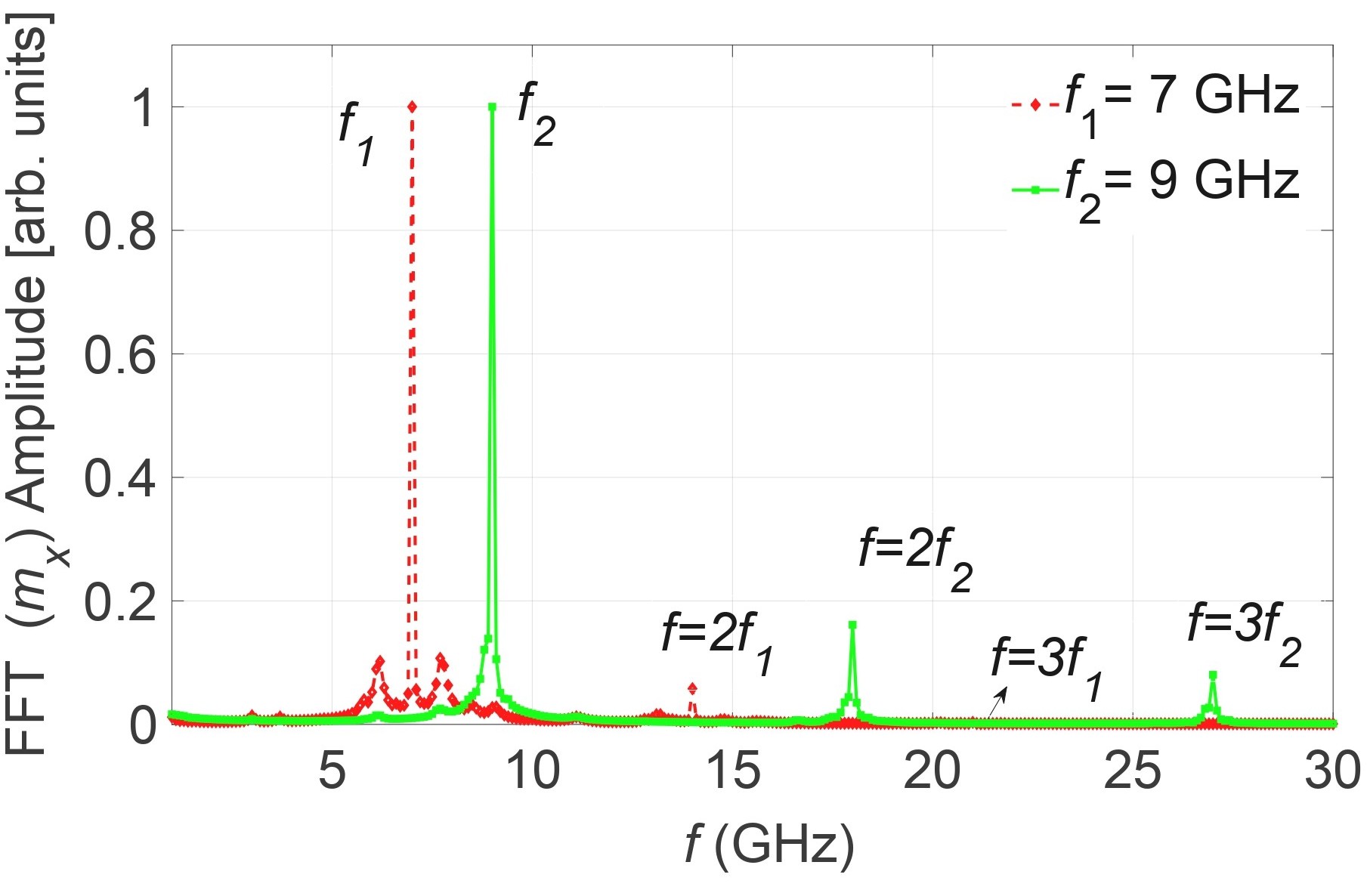}
	\caption{An illustrative example of frequency multiplication by Hopfion configuration. Under the out-of-plane excitation with field of frequency $f_1=7$ GHz and $f_2=9$ GHz {and} amplitude $2$ mT. Close-by small features near 7 GHz are nonlinear sidebands (finite-amplitude modulation / mode coupling) and they disappear in the small-drive limit }    
	\label{fig3}
\end{figure}

\begin{figure*}[hbt!]
	\centering
	\includegraphics[width=\linewidth]{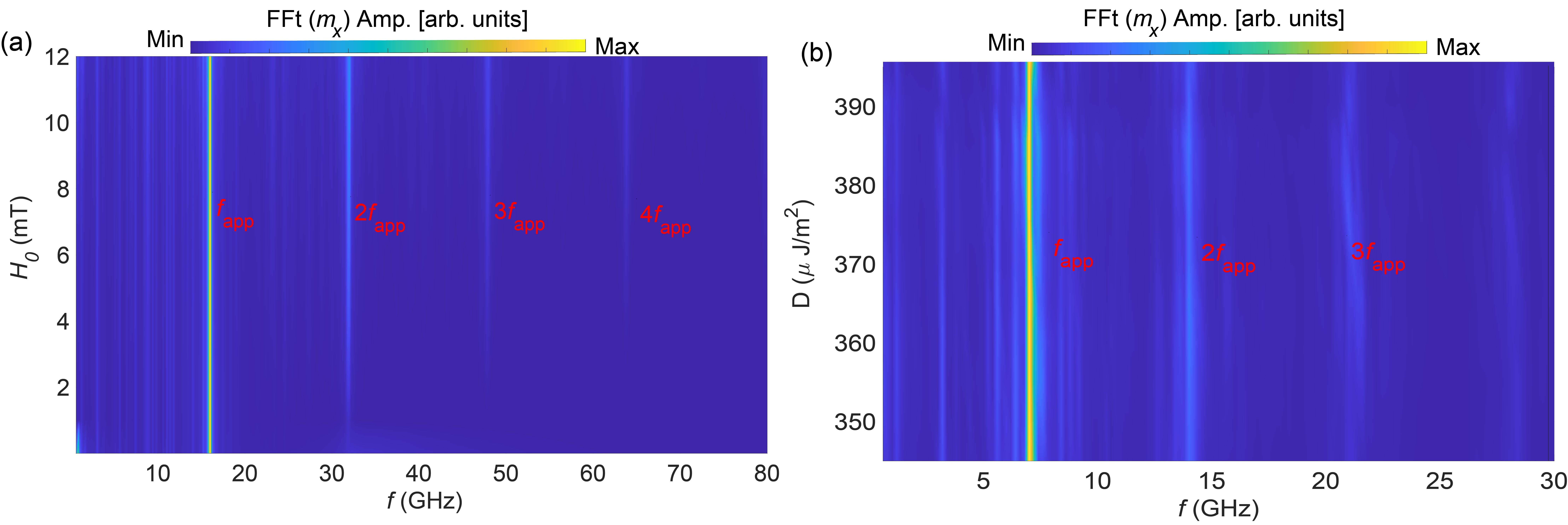}
	\caption{
\textcolor{red}{(a) Field-frequency map of the magnetization response under an in–plane microwave excitation $\mathbf{B}_{\mathrm{ext}}(t)=H_0\sin(2\pi f_{\mathrm{app}} t)\,\hat{x}$. Here applied frequecny is $f_{\mathrm{app}}=16$ GHz. The color scale shows the normalized FFT amplitude of the magnetization as a function of the frequency $f$ and the microwave field amplitude $H_0$. 
(b) DMI–frequency phase diagram of the magnetization response under a microwave excitation with frequency $f_{\mathrm{app}}=7$ GHz. The color scale shows the normalized FFT amplitude of the spatially
averaged $m_x$ component as a function of frequency and DMI strength. Clear vertical features appear at the fundamental frequency $f_{\mathrm{app}}$ and its higher harmonics indicate nonlinear frequency multiplication. The harmonic peaks persist over a wide range of driving amplitudes and DMI values, demonstrating the robustness of the nonlinear magnetization dynamics.}} 
	\label{fig3a}
\end{figure*} 
\begin{figure}[h]
	\centering

		\includegraphics[width=\linewidth]{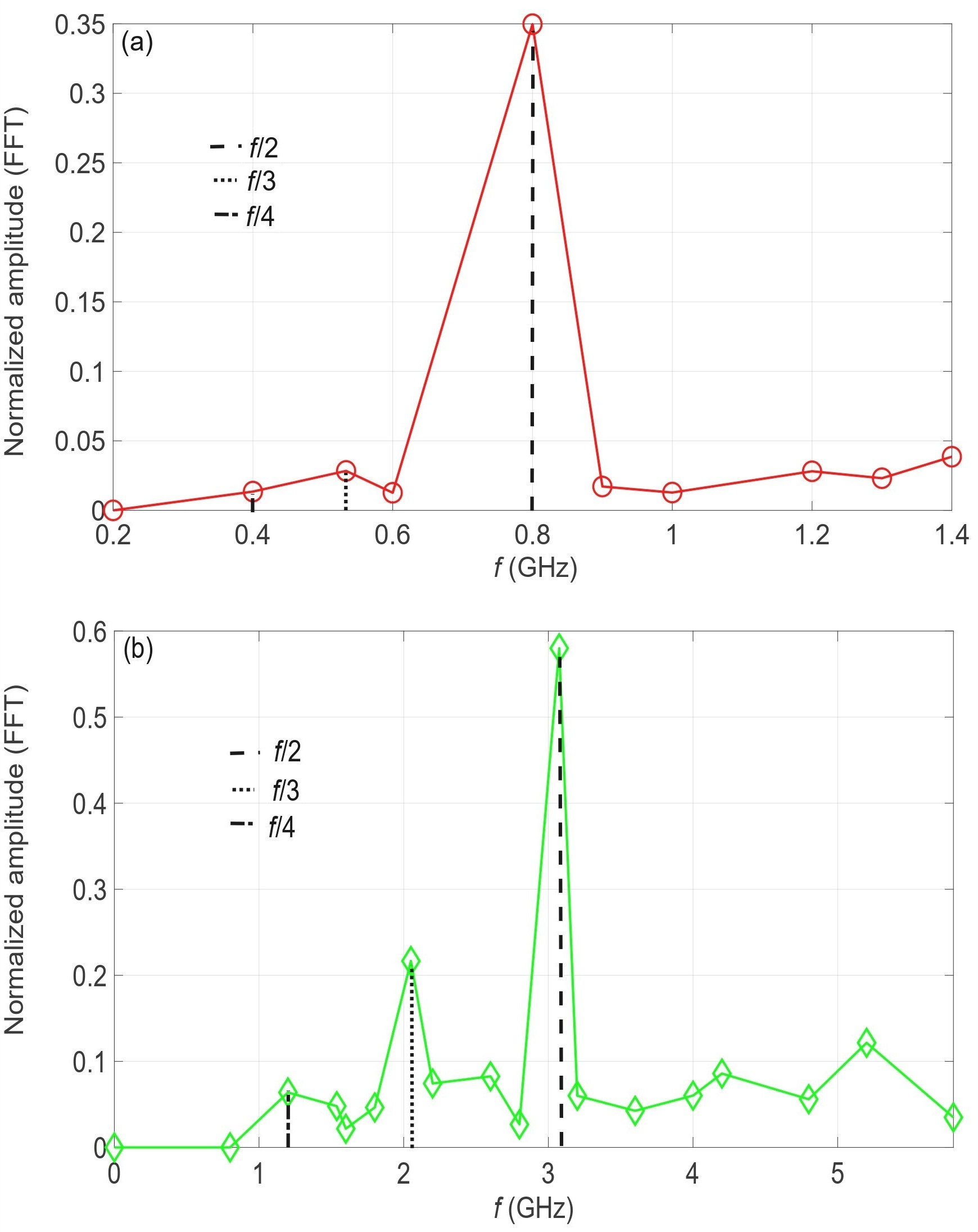}
	\caption{Illustrative examples of eigenmode excitation by microwave with fractional frequency. (a) In-plane excitation of eigenmode by applied frequencies that are fraction  of localized mode  $f=1.6$ GHz.  (b) Out-of-plane excitation of eigenmode by applied frequencies that are fraction of localized mode at $f=6.15$ GHz. In both examples we have normalized the recorded amplitude of FFT at the corresponding eigenfrequencies by dividing the amplitude of applied frequency, i.e., FFT amplitude at 1.6 GHz, by the FFT amplitude at applied frequency. }  
\label{fig4}
	\label{fig:combined}
\end{figure}
\section{\label{sec:level2} Simulation details}
We consider a circular disk characterized by a diameter $d=200$ nm and height $h=90$ nm. The schematic diagram of {setup} is shown in  {Fig. \ref{fig1}(a)}. The circular disk is a chiral ferromagnet with vanishing crystalline anistropy and sandwiched by two layers with a high perpendicular magnetic anisotropy (PMA) of thickness $2$ nm. 
 Such kind of PMA boundaries are necessary to stabilize the Hopfion \cite{liu2018binding,sutcliffe2018hopfions,tai2018static}. The {energy density} for this system can be written as
%\begin{equation}
%	E=A(\nabla\cdot \boldsymbol{m})^2 + %(1-q){D}\boldsymbol{m}\cdot{\nabla\times\boldsymbol{m}}+q %K_u (m_z)^2 -\frac{1}{2} Ms \boldsymbol{B}_d\cdot\boldsymbol{m},
%	\label{eq1}
%\end{equation}
\begin{equation}
	\begin{split}
		E = & A(\nabla\boldsymbol{m})^2 + (1-q){D}\boldsymbol{m}\cdot(\nabla\times\boldsymbol{m}) \\
		& + q K_u (m_z)^2 - Ms \boldsymbol{B}\cdot\boldsymbol{m}+E_\text{demag},
	\end{split}
	\label{eq1}
\end{equation}

%\begin{figure}[hbt!]
%	\centering
%	\includegraphics[width=\linewidth]{image/DMI_vs_field.pdf}
%	\caption{
%\textcolor{red}{DMI–frequency phase diagram of the magnetization response under a
%microwave excitation with frequency $f_{\mathrm{app}}=7$ GHz.
%The color scale shows the normalized FFT amplitude of the spatially
%averaged $m_x$ component as a %function of frequency and DMI strength.
%A strong response appears at the %driving frequency $f_{\mathrm{app}}$
%nd higher harmonics at $2f_{\mathrm{app}}$, $3f_{\mathrm{app}}$, demonstrating nonlinear frequency multiplication.} }
%	\label{fig3b}
%\end{figure} 

\noindent where $A$ is the exchange constant, $\boldsymbol{m}$ is the unit magnetization vector, ${D}$ is the Dzyaloshinskii-Moriya interaction (DMI), $q$ is a parameter used to define two different regions, the value of which is 1 for the two capping layers of PMA and zero for the central disk, $K_u$ is uniaxial anisotropy coefficient, $M_s$ is saturation magnetization, $\boldsymbol{B}$ is the external magnetic field, and $E_\text{demag}$ is the  dipolar interaction. To simulate the  magnetization dynamics, we use MuMax3 that numerically solves  the LLG equation,
\begin{equation}
	\frac{ \partial \boldsymbol{m}}{ \partial t}=-\gamma \boldsymbol{m} \times \boldsymbol{H}_{{eff}}+\alpha \boldsymbol{m}\times \frac{ \partial \boldsymbol{m}}{ \partial t},
	\label{eq3}
\end{equation}
where $\boldsymbol{H}_{{eff}}=-\frac{1}{\mu_0 M_s}\frac{\delta E}{\delta \boldsymbol{M}}$ is the effective field. Throughout this paper, the following parameters (unless otherwise stated) are chosen to solve Eq. \eqref{eq3},  $M_s=384$ kA/m,  $A=2.19$ pJ/m,  $340<D<395$ $\mu$J/m$^2$,   $K_u=8\times10^5$J/m$^3$ and $\alpha=0.01$ with the cell size being $2\times2\times2\text{ nm}^3$. Several initial magnetization configurations have been shown to evolve into Hopfions. For example, setting $\boldsymbol{m}=(0,0,\pm1)$ in the top and bottom regions, while the interior region contains a Bloch-type skyrmion structure \cite{liu2018binding}. We adopt the following ansatz as the initial condition \cite{sutcliffe2018hopfions} %in the cylindrical coordinates $(\rho,\phi,z)$
\begin{equation}
	\boldsymbol{m}=\left[\sin(2 \pi \rho/z) \sin\phi,-\sin(2 \pi \rho/z)\cos \phi,\cos(2 \pi \rho/z)\right],
	\label{eq2}
\end{equation}
where $\rho=\sqrt{x^2+y^2}$, $\phi=\arctan(y/x)$. Following { Ref.} \cite{joos2023tutorial}, we first relax the system, and then run the simulation for 2 ns to achieve a stable Hopfion.  The resulting ground state in the chiral nanodisk is illustrated in Fig. \ref{fig1}. We have checked that the Hopfion is stable for a range of material parameters, i.e., $340<D<395$ $\mu$J/m$^2$. 

To obtain the spin-wave resonance spectrum of the Hopfion, we perturbed the equilibrium magnetization state with field of the form $\mathbf{B}_{\mathrm{ext}}(t)=H_{\text{0}}\,\text{sinc}\left[2 \pi f_{c} (t-t_0)\right]\hat{e}$ where $\hat{e}=\{\hat{x},\hat{z}\}$ is the applied field direction, $f_{c}$ is the cutoff frequency, $t_0$ is offset in the sinc pulse, and $H_{\text{0}}$ is the applied field strength. We only excite the central region with {$t_0=1$} ns, $f_{c}=$ 50 GHz and $H_{\text{0}} = $ 5 mT. The simulation was run for 40 ns {with fix\_dt=0} and data was recorded after every 9.53 ps. Corresponding magnon resonance spectra for in-plane $\hat{e}=\hat{x}$ excitation is illustrated in Fig. \ref{fig2}. It was noted that an out-of-plane excitation $\hat{e}=\hat{z}$ excitation  generates similar propagating spin-wave resonance spectrum.  However, the localized eigenmodes excited by the two drive geometries differ qualitatively. 
In-plane excitation (along $\hat{x}$) predominantly couples to azimuthal, \emph{gyrotropic} modes: the Hopfion's inner and outer rings exhibit coherent azimuthal motion (clockwise or counterclockwise), characteristic of gyration. 
By contrast, out-of-plane excitation (along $\hat{z}$) preferentially excites axisymmetric \emph{breathing} modes in which the rings undergo radial expansion and contraction, with distinct phase relationships between the inner and outer rings. Representative spatial mode profiles for excitation along $\hat{x}$ at $f=1.6\ \mathrm{GHz}$ and along $\hat{z}$ at $f=6.1\ \mathrm{GHz}$ are shown in \cite{link}. The effect of DMI on the magnon frequency is also shown in the inset of Fig. \ref{fig2}. \textcolor{red}{Eigenmodes are labeled, and their corresponding spectra as a function of the DMI strength are plotted. Increasing the DMI enlarges both the inner and outer ring radii of the Hopfion, while having only a minor influence on the magnon eigenfrequencies. This indicates that the localized magnon modes are primarily determined by the overall Hopfion topology rather than small variations in the DMI strength.}

\section{\label{sec:level2} Nonlinear Dynamics}

A strong perturbation to Hopfion can induce a nonlinear magnon interaction due to the nonuniform magnetization. At high magnon densities, one may expect many-magnon scattering and mode coupling. In Fig. \ref{fig3}, we demonstrate that the noncollinear spin texture inside the Hopfion enhances the magnon-magnon coupling which can lead to frequency multiplication of the driven microwave field.  An out-of-plane sinusoidal function $\bold{B}_{\text{ext}}=H_{0}\sin(2 \pi f_n t)_{n=1,2} \hat{z}$ with $H_{0}=2$ \text{mT} is applied to excite the region $q=0$ with frequencies $f_1=7$ GHz and $f_2=9$ GHz.  %{The  fourier transform (FFT) of the magnetization spectrum also exhibits peaks at 6.15 and 7.7 GHz, corresponding to eigenmodes distinct from the applied frequency.} 
However, due to the noncollinear spin texture of the Hopfion configuration, both perturbations generate harmonics of the applied frequency $2f_1$, $2f_2$, $3f_1$ and  $3f_2$.

\textcolor{red}{To further characterize the nonlinear magnetization dynamics, we compute a field–frequency map of the FFT response under a microwave excitation 
$\mathbf{B}_{\mathrm{ext}}(t)=H_0\sin(2\pi f t)\,\hat{x}$. 
Figure \ref{fig3a} (a) shows the normalized FFT amplitude as a function of the frequency and the excitation amplitude $H_0$ with $D=395$ $\mu$J/m$^2$ and $f_{\mathrm{app}}=16$ GHz.  Distinct vertical bands appear at the fundamental excitation frequency $f_{\mathrm{app}}$ and its higher harmonics $2f_{\mathrm{app}}$, $3f_{\mathrm{app}}$, and $4f_{\mathrm{app}}$. These features indicate nonlinear frequency multiplication arising from the magnetization dynamics. Importantly, the harmonic response persists across a broad range of microwave field amplitudes, indicating that the nonlinear excitation mechanism is robust and not restricted to a narrow parameter regime. Such behavior is characteristic of strongly nonlinear magnetic textures, where mode coupling and nonlinear precession lead to efficient harmonic generation.}

\textcolor{red}{To examine the role of the DMI, we compute a DMI-frequency phase diagram for a microwave excitation 
$\mathbf{B}_{\mathrm{ext}}(t)=H_0\sin(2\pi f t)\,\hat{x}$, for $f=7$ GHz. Normalized FFT amplitude as a function of the frequency $f$ and DMI is shown in Fig. \ref{fig3a} (b). The harmonic features remain largely vertical across the investigated DMI range, indicating that their frequencies are set by the external drive. However, a slight broadening of the spectral features (i.e., in 3rd harmonics) is observed. This likely reflects increased mode coupling and reduced dynamical coherence as the Hopfion approaches the limits of its stability regime.}

\begin{figure}[hbt!]
	\centering
\includegraphics[width=\linewidth]{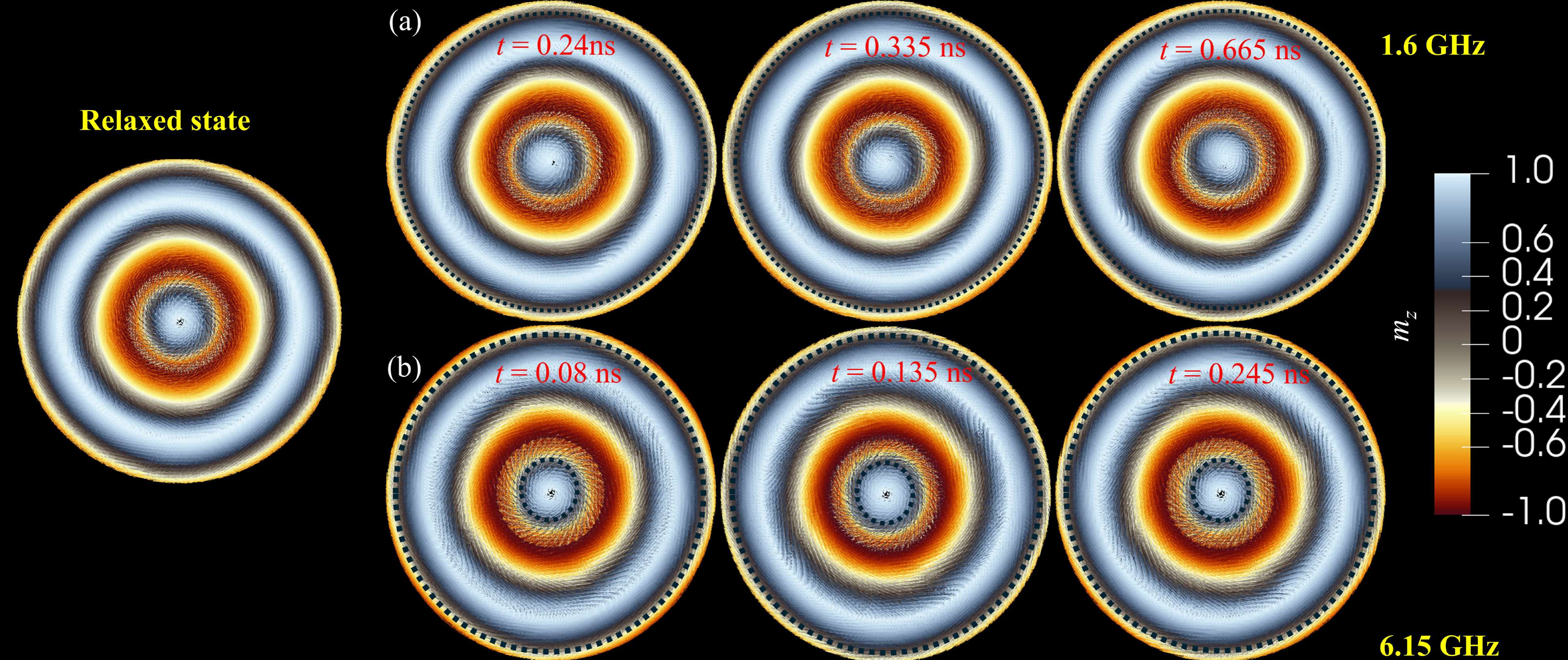}
\caption{{(a) Spatial profile ($z=0$) of the localized mode are shown at various time slots for $f=1.6$ GHz note that the inner region circulates in clockwise manner while the the outer ring vibrate in a circle drawn. (b) Spatial profile of the localize mode are shown at different time slots for $f=6.15$ GHz where the inner and outer rings expand and contract (two black circle are also drawn for comparison.)}. For enhanced visualization, animated GIFs are provided (see Reference~\cite{link}).}
	\label{fig4a}
\end{figure}

{In Fig. \ref{fig2}, a number of localized magnon states of the Hopfion are labeled in the spin-wave resonance spectrum. Harmonics generated by a driven perturbation at a fractional frequency of localized magnon state can induce resonance, thereby exciting the target localized magnon states.} No feasible threshold is observed for the driving field, indicating that the frequency multiplication can occur with very small perturbation. The efficiency of this nonlinear process is enhanced by increasing the driving amplitude \cite{rodrigues2021nonlinear}. To study the resonance with {in/out-of-plane drivings} that are fractions of eigenfrequency of the magnon localized state, we calculate the amplitude of eigen mode as a function of applied frequency. Figure \ref{fig4} (a) {demonstrates in-plane excitation} for eigenmodes at $f=1.6$ GHz. These peaks at applied frequencies $f/2$,$f/3$ and $f/4$ {indicate} that driven perturbation can {effectively} excite the target magnon localized states. In Fig. \ref{fig4} (b), out-of-plane excitation for $f=6.15$ GHz is represented graphically. A large amplitude is observed for driving frequencies of $f/2$, $f/3$ and $f/4$. {Spatial profile of these localized modes at different time slot is shown in Fig. \ref{fig4a}. Spatial profiles of the localized mode are shown in Fig. \ref{fig4a} (a) at various time intervals for a frequency of $f$=1.6 GHz. The inner region circulates in a clockwise direction, while the outer ring oscillates along the drawn circular boundary. Spatial profiles of the localized mode are shown in \ref{fig4a} (b) at different time intervals for a frequency of $f$=6.15 GHz. The inner and outer rings exhibit expansion and contraction, with two black circles included for comparison. For enhanced visualization, see animation in Ref. ~\cite{link}.}

\begin{figure*}[hbt!]
	\centering
\includegraphics[width=.9\linewidth]{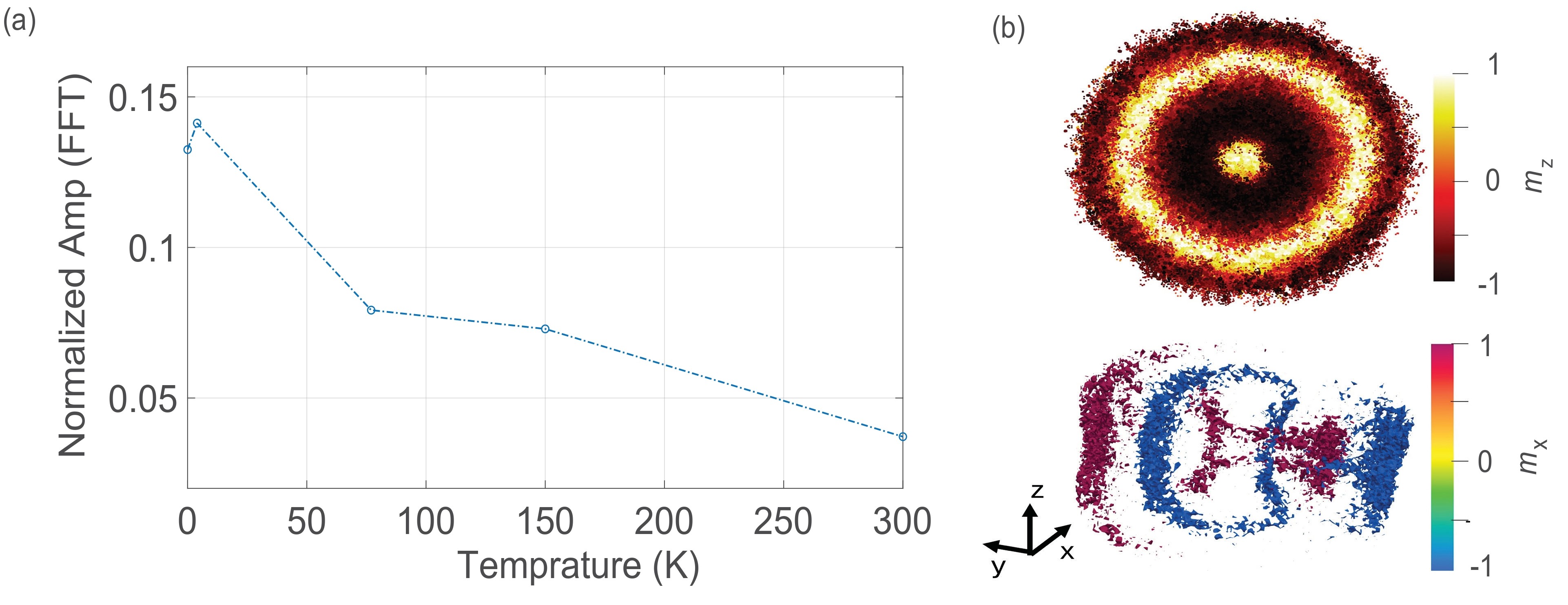}
\caption{\textcolor{red}{(a) Illustrates the temperature dependence of the normalized second-harmonic (2f) amplitude of the Hopfion localized mode. The system is driven at $f_{\mathrm{app}}=6.15/2$ GHz to excite mode at 6.15 GHz.  (b) Illustrates the  effect of on hopfion profile and isosurface at $\textbf{m}=\left(\pm1,0,0\right)$ for 300 K. }}
	\label{fig4ab}
\end{figure*}

\textcolor{red}{To evaluate thermal robustness, we computed the second-harmonic response under stochastic LLG dynamics from 0 K to 300 K. As shown in Fig. \ref{fig4ab} (a), the normalized 2f amplitude decreases monotonically with increasing temperature, dropping by approximately 70\% at 300 K compared to 0 K. This suppression arises from thermal phase decoherence, linewidth broadening, and reduced nonlinear coupling efficiency. Importantly, the second harmonic remains finite at room temperature, indicating that frequency multiplication via Hopfion dynamics is robust against realistic thermal fluctuations. Similarly the effect of temperature at 300 K on hopfion profile is illustrated in Fig. \ref{fig4ab} (b). Notice the hopfion core and preimage topology remain recognizable at 300 K, indicating structural persistence under realistic thermal fluctuations.}

\textcolor{red}{To investigate the effect of material inhomogeneity on the nonlinear dynamics, we introduce quenched disorder in the DMI. A polycrystalline grain structure is generated using the \texttt{ext\_makegrains} routine in MuMax3 with an average grain size of approximately $2\,\mathrm{nm}$. Each grain is assigned a randomized DMI value of the form $D_i = D_0 + \delta D_i$, where $\delta D_i$ is drawn from a normal distribution with a standard deviation of $0.1D_0$. In addition, the exchange coupling between neighboring grains is reduced by $10\%$ to model weakened grain boundaries.}

\begin{figure}[hbt!]
	\centering
\includegraphics[width=\linewidth]{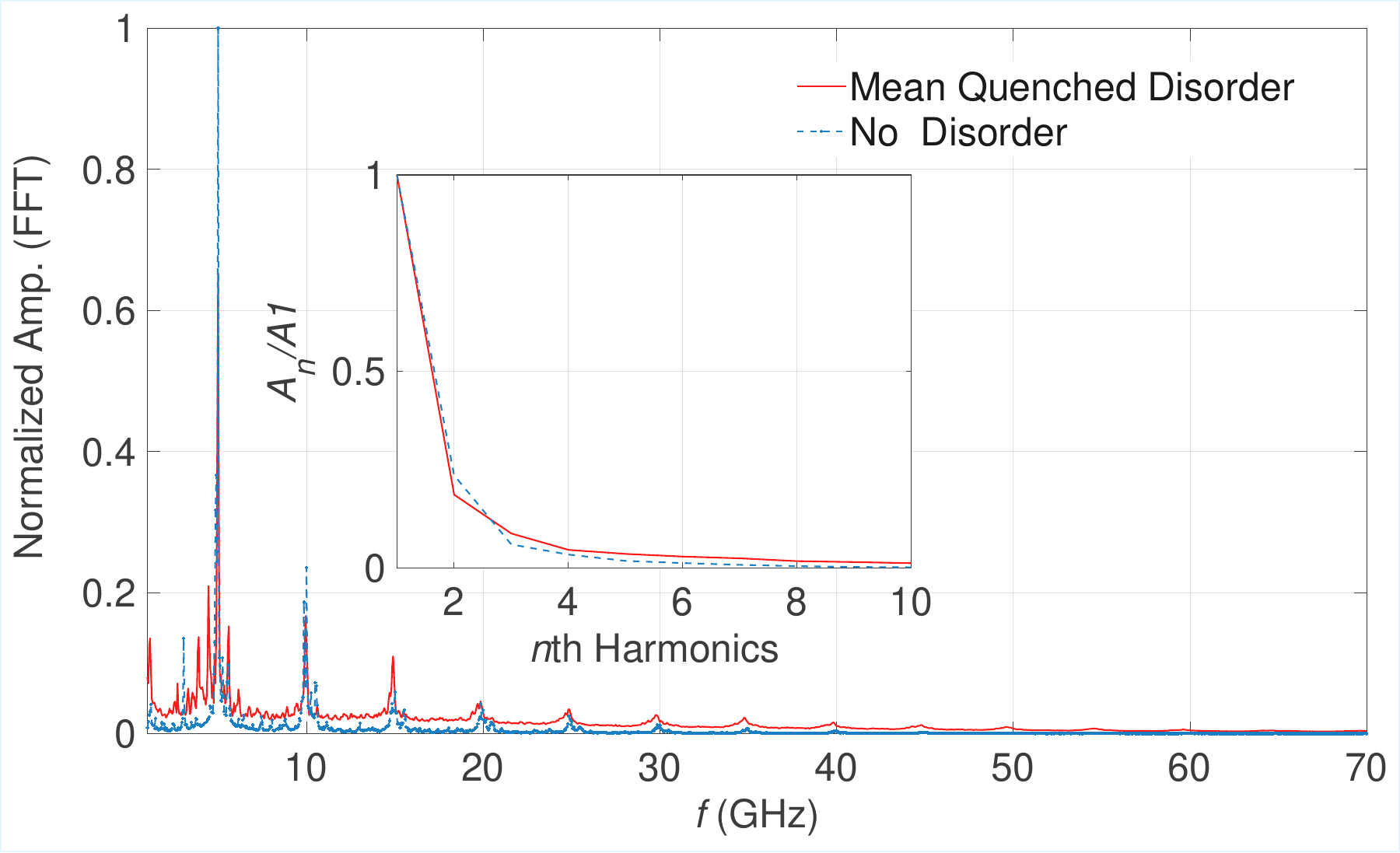}
\caption{
\textcolor{red}{Mean FFT spectrum of the magnetization response under microwave excitation for uniform and quenched-DMI systems is illustrated. The dashed blue curve corresponds to a uniform sample with $D=D_0$, while the solid red curve shows the mean over four quenched-disorder realizations where the DMI is randomized per grain ($\sigma=0.1D_0$). The grain structure is generated using \texttt{ext\_makegrains} (grain size $\approx 2\,\mathrm{nm}$) with a $10\%$ reduction of inter-grain exchange. Spectra represent the mean FFT amplitude averaged over four random seeds and normalized to the fundamental peak. The inset shows the harmonic envelope $A_n/A_1$ for $n=1\ldots10$. Quenched disorder slightly broadens the peaks and enhances higher-harmonic amplitudes.}
}
    \label{fig4ac}
\end{figure}

\textcolor{red}{To obtain statistically reliable results, four independent disorder realizations with different random seeds are simulated. The system is driven by an in-plane microwave magnetic field  $\mathbf{B}_{\mathrm{ext}}(t)=H_0\sin(2\pi f_{\mathrm{app}} t)\,\hat{x}$ with 
$f_{\mathrm{app}}=5\,\mathrm{GHz}$ and $H_0=3\,\mathrm{mT}$. The resulting FFT spectra are averaged to obtain the mean spectral response shown in Fig. \ref{fig4ac}. Compared to the uniform system, the presence of quenched DMI disorder leads to a modest broadening of the resonance peaks and a systematic increase in the higher harmonic amplitudes. This behavior indicates that spatial variations in DMI enhance nonlinear mode mixing and frequency multiplication while preserving the overall harmonic structure of the excitation spectrum.}


\begin{figure*}[hbt!]
	\centering
\includegraphics[width=\linewidth]{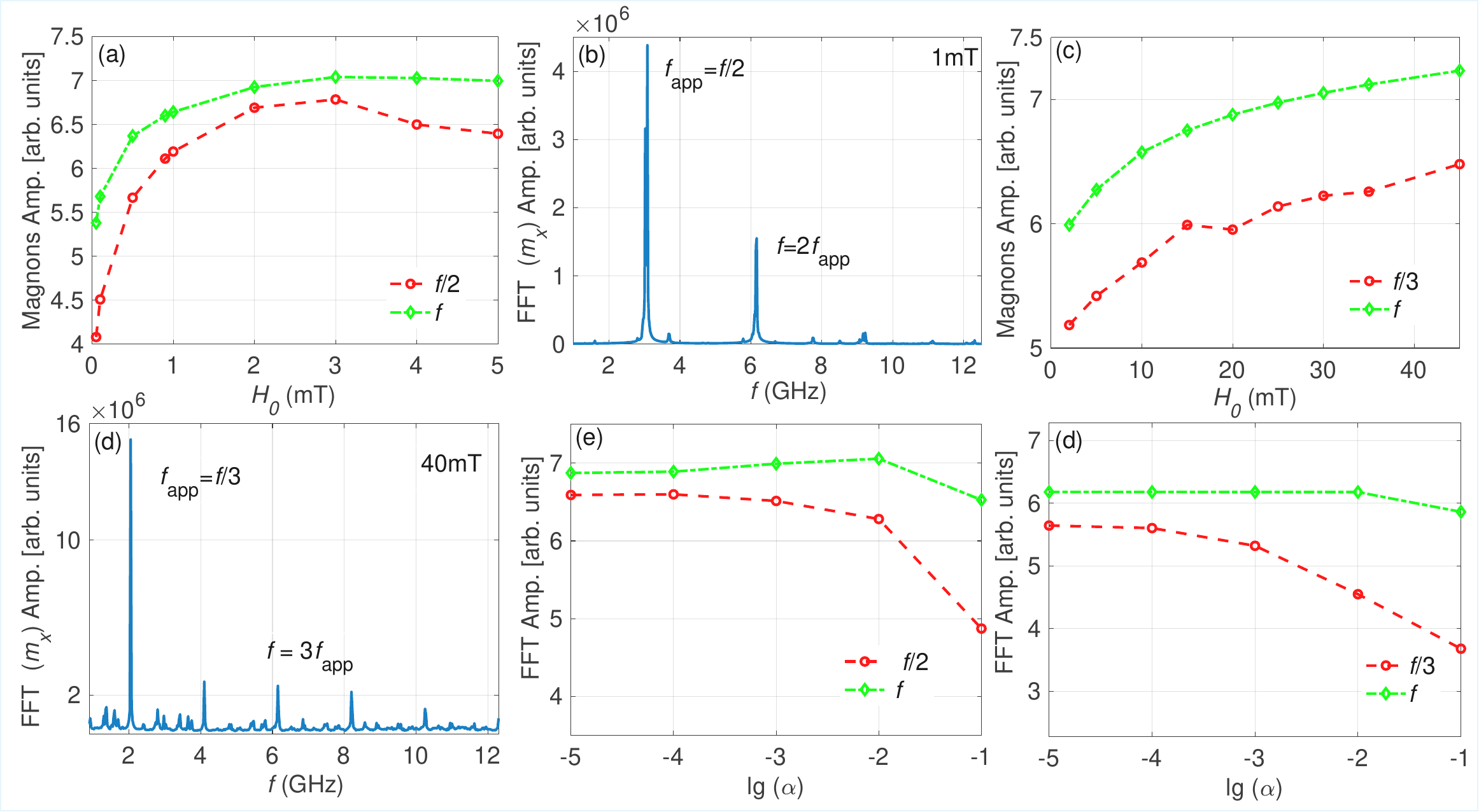}
	\caption{(a) Effect on localized mode amplitude as function of applied field strength.
		(b) Illustration of the excitation of the localized mode by 2nd harmonics of applied field.
		(c) Impact of an applied field strength on the localized mode amplitude.
		(d) Illustration of the excitation of the localized mode by a 3rd harmonics of applied field..
		(e) Effect of varying $\alpha$ on localized mode amplitude for an applied frequency of 1/2 $f$.
		(f) Effect of varying  $\alpha$ on localized mode amplitude for an applied frequency of 1/3 $f$
		. }
	\label{fig5}
\end{figure*}

\begin{figure}[hbt!]
	\centering
\includegraphics[width=\linewidth]{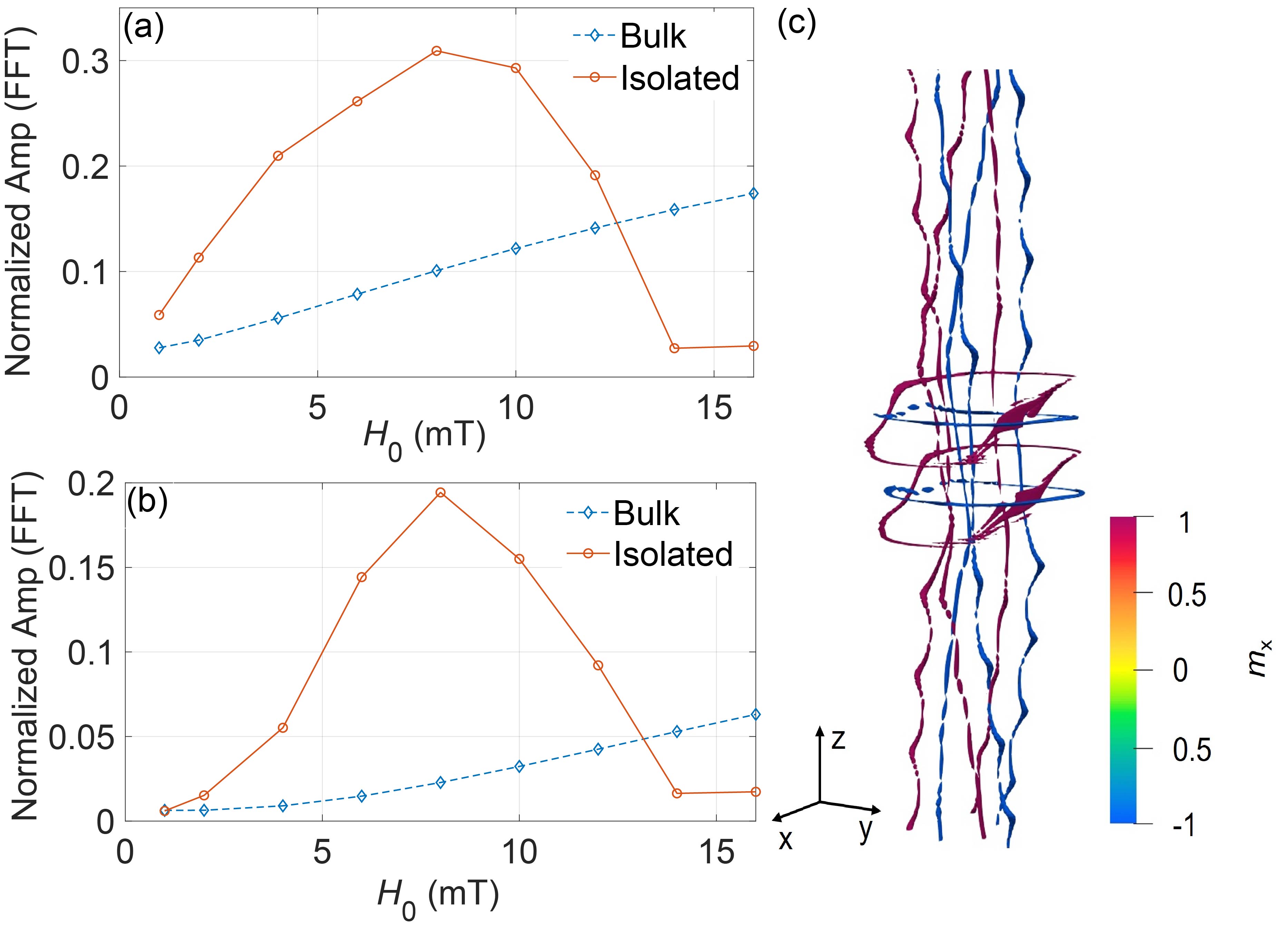}
	\caption{\textcolor{red}{Comparison of isolated and bulk Hopfion harmonic response. 
	(a) Amplitude of the second harmonic (2$f$) and (b) amplitude of the third harmonic (3$f$) plotted as a function of in-plane microwave with frequency $f_{\mathrm{app}} = 7$\,GHz. Solid lines: isolated Hopfion; dashed lines: bulk Hopfion assembly (simulation parameters given in Sec.~II). (c) $\mathbf{m}=\{\pm1,0,0\}$ isosurface and mid-plane map from the bulk simulation showing multiple Hopfion rings and linked skyrmion strings. }}
	\label{fig:hopfion_vs_bulk}
\end{figure}

In Fig. \ref{fig5} the effect of the applied magnetic field strength on the harmonics of out-of-plane driven perturbation at $f/2$ and $f/3$ for the target mode $f=6.15$ GHz is presented.  It was observed that a driven perturbation at half the eigen frequency $f$ can still excite the eigen mode, even with an applied field with the strength as low as $0.05$ mT.  {In contrast, the perturbation at 
$f/3$ relies on third-order nonlinearity, which is weak. As a result, the energy transfer to the eigenmode is insufficient, requiring much higher  excitation field strengths.}
%\begin{figure*}[hbt!]
	%\centering
	%\begin{tabular}{c c}
%		\includegraphics[width=0.455\linewidth,height=.5655\columnwidth]{image/a1a} & 
%		\includegraphics[width=0.455\linewidth,height=.59\columnwidth]{image/b1a} \\
%		\includegraphics[width=0.455\linewidth,height=.565\columnwidth]{image/c1a} & 
%		\includegraphics[width=0.455\linewidth,height=.59\columnwidth]{image/d1a} \\
%		\includegraphics[width=0.455\linewidth,height=.565\columnwidth]{image/e1a} & 
%		\includegraphics[width=0.455\linewidth,height=.565\columnwidth]{image/f1a} \\
%	\end{tabular}
%	\caption{(a) Effect on localized mode amplitude as function of applied field strength.
%		(b) Illustration of the excitation of the localized mode by 2nd harmonics of applied field.
%		(c) Impact of an applied field strength on the localized mode amplitude.
%		(d) Illustration of the excitation of the localized mode by a 3rd harmonics of applied field..
%		(e) Effect of varying $\alpha$ on localized mode amplitude for an applied frequency of 1/2 $f$.
%		(f) Effect of varying  $\alpha$ on localized mode amplitude for an applied frequency of 1/3 $f$
	%	. }
%	\label{fig5}
%\end{figure*}

It is also evident that as the strength of the applied magnetic field increases, both the FFT amplitudes of the  $f/2$ and $f$ components increase in a nonlinear manner as shown in Fig. \ref{fig5}(a). The rate of change of the amplitude is high at first and slow after 3 mT. {This likely reflects saturation effects or nonlinear damping mechanisms that limit further growth of the magnon number.} The corresponding FFT  spectrum for the applied field strength of $1$ mT is plotted in Fig. \ref{fig5}(b). In the case of driven perturbation, $f/3$, its harmonics increase with the increase of field strength as shown in Fig. \ref{fig5}(c). The FFT amplitude spectrum for the case $40$ mT is also illustrated in Fig. \ref{fig5}(d). In Figs. \ref{fig5}(c) and \ref{fig5}(f) effect of $\alpha$ on the FFT amplitude is shown.  We see that for both forced perturbation, i.e., $f/3$ and $f/2$, the FFT amplitude decreases with the increase of $\alpha$ suggesting { that a higher damping coefficient will reduce the efficiency of energy transfer into the targeted eigenmode. In addition, amplitude drop for $f/3$ is more rapid than $f/2$ due to its third-order nonlinear nature.  }

\textcolor{red}{So far, we have studied the nonlinear response of an isolated Hopfion, which is a localized topological texture confined in a finite nanostructure, to a microwave field. It is therefore interesting to examine the corresponding response in a bulk Hopfion, which consists of multiple linked or extended Hopfion-like textures embedded in a three-dimensional magnetic volume. Figure~\ref{fig:hopfion_vs_bulk} compares the nonlinear response of an isolated Hopfion and a bulk assembly computed using their respective parameter sets (see Sec.~II). For the bulk case, simulations were performed with $M_s = 3.84 \times 10^5~\mathrm{A\,m^{-1}}$, $A_{\mathrm{ex}} = 4.75 \times 10^{-12}~\mathrm{J\,m^{-1}}$, and a characteristic length $L_D = 70~\mathrm{nm}$ (yielding $D = 4\pi A_{\mathrm{ex}}/L_D$). The simulation volume was $5L_D \times 5L_D \times 10L_D$, discretized into $128 \times 128 \times 256$ cells with magnetostatic fields enabled. After relaxation, the system was driven in-plane with a sinusoidal field at $f = 7~\mathrm{GHz}$ for $4~\mathrm{ns}$.} \textcolor{red}{At low-to-moderate drive amplitudes, the isolated Hopfion produces the largest $2f$ and $3f$ harmonic amplitudes, reflecting its strongly localized three-dimensional magnetization texture and a small number of well-defined localized magnon eigenmodes that couple efficiently to the drive. When the drive exceeds the threshold, these localized modes are disrupted and the harmonic amplitudes decrease sharply. By contrast, the bulk system shows a steadier and more monotonic increase in the total harmonic response with drive amplitude. In this case, the injected energy is redistributed across a larger volume and among coupled topological units, resulting in a more robust but lower response from each individual Hopfion or localized region.}
\textcolor{red}{Taken together, Fig. \ref{fig:hopfion_vs_bulk} reveals a clear trade-off: an isolated Hopfion offers stronger harmonic generation from a single localized object at practical drive levels, whereas the bulk system provides greater robustness and smoother scaling with drive amplitude. This trade-off is expected to become more pronounced for larger Hopf index $H$  (the topological invariant that counts the linking number of the preimage loops), where the increased number of internal modes and stronger coupling between localized units broaden the excitation spectrum and reduce the harmonic response of an individual object unless additional stabilization mechanisms, such as increased thickness, larger $D$, confinement, or multilayer engineering, are introduced.}

\section{\label{sec:level5}Summary}

In this work, we investigated the nonlinear dynamics of magnetic Hopfions. We showed that the Hopfion’s ring structure expands with increasing Dzyaloshinskii–Moriya interaction (DMI) strength, reflecting the system’s tendency to minimize its total energy by distributing the DMI-favored spin twist over a larger region and thereby alleviating the exchange penalty. We further analyzed the magnon spectrum under both in-plane and out-of-plane excitations, identifying discrete magnon states localized within the Hopfion’s topological texture. Forced perturbations were found to generate higher harmonics of the drive frequency, demonstrating intrinsic nonlinear responses. Importantly, we showed that localized magnon states can be efficiently excited when the system is driven at fractional multiples of their eigenfrequencies, enabling frequency multiplication mediated by the Hopfion. Since these localized modes decay rapidly away from the texture, fractional excitation provides a practical pathway for selective mode addressing at accessible field amplitudes, which is highly relevant for experimental detection.

\textcolor{red}{Building on the original results, we quantified the robustness of these nonlinear effects against thermal fluctuations and quenched material inhomogeneity. Stochastic-LLG simulations show that the Hopfion core and preimage topology remain recognizable up to room temperature, although the normalized second-harmonic amplitude is progressively suppressed (the $2f$ signal falls notably with $T$, by on the order of tens of percent up to $300\,$K in our simulations) due to thermal decoherence and effective linewidth broadening. Likewise, introducing quenched DMI disorder on a polycrystalline grain map (mean grain size $\approx2\,$nm, $\sigma_{D}\approx0.1D_0$) modestly broadens resonance peaks and, interestingly, enhances the relative weight of higher harmonics via disorder-induced mode mixing while preserving the overall harmonic structure.}
\textcolor{red}{Finally, we compared isolated Hopfion behavior with that of a bulk assembly. The isolated Hopfion provides the largest per-object $2f$ and $3f$ amplitudes at low-to-moderate drives because of its strongly localized eigenmodes, but these amplitudes collapse above a deformation threshold as localized modes are disrupted. By contrast, coupled bulk assemblies redistribute injected energy across many rings and longitudinal channels, yielding steadier, more monotonic ensemble harmonic growth with drive at the expense of reduced per-ring conversion efficiency. Taken together, these extended results deepen our understanding of Hopfion nonlinear dynamics and strengthen the case for Hopfion-based, low-power magnonic frequency multipliers and signal processors, while also identifying the key stability and materials considerations (temperature, disorder, and coupling) that must be addressed for device implementation.}

\section*{acknowledgment}
This project was supported by the National Key R\&D Program China (Grant No. 2022YFA1402802), National Natural Science Foundation of China (Grant No. 12374103 and 12434003), and Sichuan Science and Technology Program (No. 2025NSFJQ0045) 
\section*{Code availability }
The codes for Hopfion generation and microwave excitation are available in a public repository.  
https://github.com/waleedwaseer/Hopfion

\bibliography{biblo}% Produces the bibliography via BibTeX.

@book{manton2004topological,
  title={Topological solitons},
  author={Manton, Nicholas and Sutcliffe, Paul},
  year={2004},
  publisher={Cambridge University Press}
}

@article{nitta2022relations,
  title={Relations among topological solitons},
  author={Nitta, Muneto},
  journal={Physical Review D},
  volume={105},
  number={10},
  pages={105006},
  year={2022},
  publisher={APS}
}

@article{rodrigues2021nonlinear,
  title={Nonlinear dynamics of topological ferromagnetic textures for frequency multiplication},
  author={Rodrigues, Davi R and Nothhelfer, Jonas and Mohseni, Morteza and Knapman, Ross and Pirro, Philipp and Everschor-Sitte, Karin},
  journal={Physical Review Applied},
  volume={16},
  number={1},
  pages={014020},
  year={2021},
  publisher={APS}
}

@article{skyrme1962unified,
	title={A unified field theory of mesons and baryons},
	author={Skyrme, Tony Hilton Royle},
	journal={Nuclear Physics},
	volume={31},
	pages={556--569},
	year={1962},
	publisher={Elsevier}
}

@article{yu2010real,
	title={Real-space observation of a two-dimensional skyrmion crystal},
	author={Yu, XZ and Onose, Yoshinori and Kanazawa, Naoya and Park, Joung Hwan and Han, JH and Matsui, Yoshio and Nagaosa, Naoto and Tokura, Yoshinori},
	journal={Nature},
	volume={465},
	number={7300},
	pages={901--904},
	year={2010},
	publisher={Nature Publishing Group UK London}
}

@article{melcher2014chiral,
	title={Chiral skyrmions in the plane},
	author={Melcher, Christof},
	journal={Proceedings of the Royal Society A: Mathematical, Physical and Engineering Sciences},
	volume={470},
	number={2172},
	pages={20140394},
	year={2014},
	publisher={The Royal Society Publishing}
}

@article{bogdanov1989thermodynamically,
	title={Thermodynamically stable “vortices” in magnetically ordered crystals. The mixed state of magnets},
	author={Bogdanov, Alexei N and Yablonskii, DA},
	journal={Zh. Eksp. Teor. Fiz},
	volume={95},
	number={1},
	pages={178},
	year={1989}
}

@article{fert2013skyrmions,
	title={Skyrmions on the track},
	author={Fert, Albert and Cros, Vincent and Sampaio, Joao},
	journal={Nature Nanotechnology},
	volume={8},
	number={3},
	pages={152--156},
	year={2013},
	publisher={Nature Publishing Group UK London}
}

@article{kiselev2011chiral,
	title={Chiral skyrmions in thin magnetic films: new objects for magnetic storage technologies?},
	author={Kiselev, Nikolai S and Bogdanov, AN and Sch{\"a}fer, R and R{\"o}{\ss}ler, UK},
	journal={Journal of Physics D: Applied Physics},
	volume={44},
	number={39},
	pages={392001},
	year={2011},
	publisher={IOP Publishing}
}

@article{sutcliffe2018hopfions,
  title={Hopfions in chiral magnets},
  author={Sutcliffe, Paul},
  journal={Journal of Physics A: Mathematical and Theoretical},
  volume={51},
  number={37},
  pages={375401},
  year={2018},
  publisher={IOP Publishing}
}

@article{raftrey2021field,
  title={Field-driven dynamics of magnetic hopfions},
  author={Raftrey, David and Fischer, Peter},
  journal={Physical Review Letters},
  volume={127},
  number={25},
  pages={257201},
  year={2021},
  publisher={APS}
}

@article{saji2023hopfion,
  title={Hopfion-driven magnonic Hall effect and magnonic focusing},
  author={Saji, Carlos and Troncoso, Roberto E and Carvalho-Santos, Vagson L and Altbir, Dora and Nunez, Alvaro S},
  journal={Physical Review Letters},
  volume={131},
  number={16},
  pages={166702},
  year={2023},
  publisher={APS}
}

@article{wang2019current,
  title={Current-driven dynamics of magnetic hopfions},
  author={Wang, XS and Qaiumzadeh, Alireza and Brataas, Arne},
  journal={Physical Review Letters},
  volume={123},
  number={14},
  pages={147203},
  year={2019},
  publisher={APS}
}

@article{khodzhaev2022hopfion,
  title={Hopfion dynamics in chiral magnets},
  author={Khodzhaev, Zulfidin and Turgut, Emrah},
  journal={Journal of Physics: Condensed Matter},
  volume={34},
  number={22},
  pages={225805},
  year={2022},
  publisher={IOP Publishing}
}

@article{liu2018binding,
  title={Binding a hopfion in a chiral magnet nanodisk},
  author={Liu, Yizhou and Lake, Roger K and Zang, Jiadong},
  journal={Physical Review B},
  volume={98},
  number={17},
  pages={174437},
  year={2018},
  publisher={APS}
}

@article{tai2018static,
  title={Static Hopf solitons and knotted emergent fields in solid-state noncentrosymmetric magnetic nanostructures},
  author={Tai, Jung-Shen B and Smalyukh, Ivan I},
  journal={Physical Review Letters},
  volume={121},
  number={18},
  pages={187201},
  year={2018},
  publisher={APS}
}

@article{faddeev1997stable,
  title={Stable knot-like structures in classical field theory},
  author={Faddeev, Ludvig and Niemi, Antti J},
  journal={Nature},
  volume={387},
  number={6628},
  pages={58--61},
  year={1997},
  publisher={Nature Publishing Group UK London}
}

@article{ackerman2017static,
  title={Static three-dimensional topological solitons in fluid chiral ferromagnets and colloids},
  author={Ackerman, Paul J and Smalyukh, Ivan I},
  journal={Nature materials},
  volume={16},
  number={4},
  pages={426--432},
  year={2017},
  publisher={Nature Publishing Group UK London}
}

@article{kosevich1990magnetic,
  title={Magnetic solitons},
  author={Kosevich, Arnold Markovich and Ivanov, BA and Kovalev, Alexander S},
  journal={Physics Reports},
  volume={194},
  number={3-4},
  pages={117--238},
  year={1990},
  publisher={Elsevier}
}

@article{kent2021creation,
  title={Creation and observation of Hopfions in magnetic multilayer systems},
  author={Kent, Noah and Reynolds, Neal and Raftrey, David and Campbell, Ian TG and Virasawmy, Selven and Dhuey, Scott and Chopdekar, Rajesh V and Hierro-Rodriguez, Aurelio and Sorrentino, Andrea and Pereiro, Eva and others},
  journal={Nature Communications},
  volume={12},
  number={1},
  pages={1562},
  year={2021},
  publisher={Nature Publishing Group UK London}
}

@article{joos2023tutorial,
  title={Tutorial: Simulating modern magnetic material systems in mumax3},
  author={Joos, Jonas J and Bassirian, Pedram and Gypens, Pieter and Mulkers, Jeroen and Litzius, Kai and Van Waeyenberge, Bartel and Leliaert, Jonathan},
  journal={Journal of Applied Physics},
  volume={134},
  number={17},
  year={2023},
  publisher={AIP Publishing}
}

@article{kruglyak2010magnonics,
  title={Magnonics},
  author={Kruglyak, VV and Demokritov, SO and Grundler, D},
  journal={Journal of Physics D: Applied Physics},
  volume={43},
  number={26},
  pages={264001},
  year={2010},
  publisher={IOP Publishing}
}

@article{lenk2011building,
  title={The building blocks of magnonics},
  author={Lenk, Benjamin and Ulrichs, Henning and Garbs, Fabian and M{\"u}nzenberg, Markus},
  journal={Physics Reports},
  volume={507},
  number={4-5},
  pages={107--136},
  year={2011},
  publisher={Elsevier}
}

@article{gilbert2004phenomenological,
  title={A phenomenological theory of damping in ferromagnetic materials},
  author={Gilbert, Thomas L},
  journal={IEEE Transactions on Magnetics},
  volume={40},
  number={6},
  pages={3443--3449},
  year={2004},
  publisher={IEEE}
}

@article{demidov2011generation,
  title={Generation of the second harmonic by spin waves propagating in microscopic stripes},
  author={Demidov, VE and Kostylev, MP and Rott, Karsten and Krzysteczko, Patryk and Reiss, G{\"u}nter and Demokritov, SO},
  journal={Physical Review B—Condensed Matter and Materials Physics},
  volume={83},
  number={5},
  pages={054408},
  year={2011},
  publisher={APS}
}

@article{sebastian2013nonlinear,
  title={Nonlinear emission of spin-wave caustics from an edge mode of a microstructured Co 2 Mn 0.6 Fe 0.4 Si waveguide},
  author={Sebastian, T and Br{\"a}cher, T and Pirro, P and Serga, AA and Hillebrands, B and Kubota, T and Naganuma, H and Oogane, M and Ando, Y},
  journal={Physical Review Letters},
  volume={110},
  number={6},
  pages={067201},
  year={2013},
  publisher={APS}
}

@article{rousseau2014propagation,
  title={Propagation of nonlinearly generated harmonic spin waves in microscopic stripes},
  author={Rousseau, O and Yamada, M and Miura, K and Ogawa, S and Otani, Y},
  journal={Journal of Applied Physics},
  volume={115},
  number={5},
  year={2014},
  publisher={AIP Publishing}
}

@article{heinrich1985fmr,
  title={FMR linebroadening in metals due to two-magnon scattering},
  author={Heinrich, B and Cochran, JF and Hasegawa, R},
  journal={Journal of Applied Physics},
  volume={57},
  number={8},
  pages={3690--3692},
  year={1985},
  publisher={AIP Publishing}
}

@article{lenz2006two,
  title={Two-magnon scattering and viscous Gilbert damping in ultrathin ferromagnets},
  author={Lenz, K and Wende, H and Kuch, W and Baberschke, K and Nagy, K{\'a}lm{\'a}n and J{\'a}nossy, Andr{\'a}s},
  journal={Physical Review B},
  volume={73},
  number={14},
  pages={144424},
  year={2006},
  publisher={APS}
}

@article{gross2022imaging,
  title={Imaging magnonic frequency multiplication in nanostructured antidot lattices},
  author={Gro{\ss}, Felix and Weigand, Markus and Gangwar, Ajay and Werner, Matthias and Sch{\"u}tz, Gisela and Goering, Eberhard J and Back, Christian H and Gr{\"a}fe, Joachim},
  journal={Physical Review B},
  volume={106},
  number={1},
  pages={014426},
  year={2022},
  publisher={APS}
}

@article{dussaux2010large,
  title={Large microwave generation from current-driven magnetic vortex oscillators in magnetic tunnel junctions},
  author={Dussaux, A and Georges, B and Grollier, J and Cros, V and Khvalkovskiy, AV and Fukushima, A and Konoto, M and Kubota, H and Yakushiji, K and Yuasa, S and others},
  journal={Nature Communications},
  volume={1},
  number={1},
  pages={1--6},
  year={2010},
  publisher={Nature Publishing Group}
}

@article{nishimura2002magnetic,
  title={Magnetic tunnel junction device with perpendicular magnetization films for high-density magnetic random access memory},
  author={Nishimura, Naoki and Hirai, Tadahiko and Koganei, Akio and Ikeda, Takashi and Okano, Kazuhisa and Sekiguchi, Yoshinobu and Osada, Yoshiyuki},
  journal={Journal of Applied Physics},
  volume={91},
  number={8},
  pages={5246--5249},
  year={2002},
  publisher={American Institute of Physics}
}

@article{tomasello2014strategy,
  title={A strategy for the design of skyrmion racetrack memories},
  author={Tomasello, Riccardo and Martinez, E and Zivieri, Roberto and Torres, Luis and Carpentieri, Mario and Finocchio, Giovanni},
  journal={Scientific Reports},
  volume={4},
  number={1},
  pages={1--7},
  year={2014},
  publisher={Nature Publishing Group}
}

@article{carpentieri2015topological,
  title={Topological, non-topological and instanton droplets driven by spin-transfer torque in materials with perpendicular magnetic anisotropy and Dzyaloshinskii--Moriya Interaction},
  author={Carpentieri, Mario and Tomasello, Riccardo and Zivieri, Roberto and Finocchio, Giovanni},
  journal={Scientific Reports},
  volume={5},
  number={1},
  pages={16184},
  year={2015},
  publisher={Nature Publishing Group UK London}
}

@article{ruotolo2009phase,
  title={Phase-locking of magnetic vortices mediated by antivortices},
  author={Ruotolo, Antonio and Cros, V and Georges, B and Dussaux, A and Grollier, J and Deranlot, C and Guillemet, R and Bouzehouane, K and Fusil, S and Fert, A},
  journal={Nature Nanotechnology},
  volume={4},
  number={8},
  pages={528--532},
  year={2009},
  publisher={Nature Publishing Group UK London}
}

@article{fleury1968scattering,
  title={Scattering of light by one-and two-magnon excitations},
  author={Fleury, PA and Loudon, R},
  journal={Physical Review},
  volume={166},
  number={2},
  pages={514},
  year={1968},
  publisher={APS}
}

@article{hurben1998theory,
  title={Theory of two magnon scattering microwave relaxation and ferromagnetic resonance linewidth in magnetic thin films},
  author={Hurben, MJ and Patton, CE},
  journal={Journal of Applied Physics},
  volume={83},
  number={8},
  pages={4344--4365},
  year={1998},
  publisher={AIP Publishing}
}

@article{faddeev1999partially,
  title={Partially dual variables in SU (2) Yang-Mills theory},
  author={Faddeev, Ludvig and Niemi, Antti J},
  journal={Physical Review Letters},
  volume={82},
  number={8},
  pages={1624},
  year={1999},
  publisher={APS}
}

@article{cooper1999propagating,
  title={Propagating magnetic vortex rings in ferromagnets},
  author={Cooper, NR},
  journal={Physical Review Letters},
  volume={82},
  number={7},
  pages={1554},
  year={1999},
  publisher={APS}
}

@article{babaev2002dual,
  title={Dual neutral variables and knot solitons in triplet superconductors},
  author={Babaev, Egor},
  journal={Physical Review Letters},
  volume={88},
  number={17},
  pages={177002},
  year={2002},
  publisher={APS}
}

@article{babaev2002hidden,
  title={Hidden symmetry and knot solitons in a charged two-condensate Bose system},
  author={Babaev, Egor and Faddeev, Ludvig D and Niemi, Antti J},
  journal={Physical Review B},
  volume={65},
  number={10},
  pages={100512},
  year={2002},
  publisher={APS}
}

@article{kleckner2013creation,
  title={Creation and dynamics of knotted vortices},
  author={Kleckner, Dustin and Irvine, William TM},
  journal={Nature Physics},
  volume={9},
  number={4},
  pages={253--258},
  year={2013},
  publisher={Nature Publishing Group UK London}
}

@article{zakharov1975spin,
	title={Spin-wave turbulence beyond the parametric excitation threshold},
	author={Zakharov, Vladimir E and L'vov, Viktor Sergeevich and Starobinets, Samuel Solomonovich},
	journal={Soviet Physics Uspekhi},
	volume={17},
	number={6},
	pages={896},
	year={1975},
	publisher={IOP Publishing}
}

@article{bryant1988spin,
	title={Spin-wave dynamics in a ferrimagnetic sphere},
	author={Bryant, Paul H and Jeffries, Carson D and Nakamura, Katsuhiro},
	journal={Physical Review A},
	volume={38},
	number={8},
	pages={4223},
	year={1988},
	publisher={APS}
}

@article{xiao2017parametric,
	title={Parametric autoexcitation of magnetic droplet soliton perimeter modes},
	author={Xiao, D and Tiberkevich, V and Liu, YH and Liu, YW and Mohseni, SM and Chung, Sunjae and Ahlberg, M and Slavin, AN and {\AA}kerman, Johan and Zhou, Yan},
	journal={Physical Review B},
	volume={95},
	number={2},
	pages={024106},
	year={2017},
	publisher={APS}
}

@article{zakeri2007spin,
	title={Spin dynamics in ferromagnets: Gilbert damping and two-magnon scattering},
	author={Zakeri, Kh and Lindner, J and Barsukov, I and Meckenstock, R and Farle, M and Von H{\"o}rsten, U and Wende, H and Keune, W and Rocker, J and Kalarickal, SS and others},
	journal={Physical Review B},
	volume={76},
	number={10},
	pages={104416},
	year={2007},
	publisher={APS}
}

@article{zheng2023hopfion,
	title={Hopfion rings in a cubic chiral magnet},
	author={Zheng, Fengshan and Kiselev, Nikolai S and Rybakov, Filipp N and Yang, Luyan and Shi, Wen and Bl{\"u}gel, Stefan and Dunin-Borkowski, Rafal E},
	journal={Nature},
	volume={623},
	number={7988},
	pages={718--723},
	year={2023},
	publisher={Nature Publishing Group UK London}
}

@article{guslienko2024magnetic,
	title={Magnetic Hopfions: A Review},
	author={Guslienko, Konstantin},
	journal={Magnetism},
	volume={4},
	number={4},
	pages={383--399},
	year={2024},
	publisher={MDPI}
}

@online{link,
author = {},
title = {https://github.com/waleedwaseer/Hopfion}}

\end{document}